\input{aipcheck}

\documentclass[
    ,final            % use final for the camera ready runs
  ]
  {aipproc}

\layoutstyle{8x11single}

\def\th{\theta}

\def\ka{\kappa}

\def\si{\sigma}

\def\ch{\chi}

\def\to{\rightarrow}

\def\no{\nonumber}
\def\no{\nonumber}
\def\beq{\begin{eqnarray}}
\def\eeq{\end{eqnarray}}

\def\ucmt{\mbox{cm}^2}

\def\uGeV{\mbox{GeV}}
\def\uMeV{\mbox{MeV}}

\def\uGeVt{\mbox{GeV}^2}

\def\uGeVct{\mbox{GeV}/\mbox{c}^2}
\def\nuance{\textsc{nuance}}

\def\nue{\nu_e}
\def\nueb{\bar{\nu}_e}
\def\numu{\nu_\mu}
\def\numub{\bar{\nu}_\mu}

\def\elp{e^{+}}
\def\elm{e^{-}}
\def\mup{\mu^{+}}
\def\mum{\mu^{-}}

\def\pip{\pi^{+}}

\def\evki{146070}
\def\MBosc1POT{5.58\times 10^{20}}
\def\distCCQEef{26}
\def\distCCQEpu{78}

\def\NEWMAcon{1.35}
\def\NEWMAerr{0.17}
\def\NEWKAcon{1.007}
\def\NEWKAerr{0.012}
\def\NUAMAcon{1.03} 

\def\NUAKAcon{1.000}

\def\CHcon{47.0}
\def\absXScon{9.412}
\def\ttXSerr{10.8}

\begin{document}
\title{First Measurement of Muon Neutrino Charged Current Quasielastic 
(CCQE) Double Differential Cross Section}


\author{Teppei Katori for the MiniBooNE collaboration}{
  address={Massachusetts Institute of Technology, Cambridge, MA
\footnote{Ph.D thesis work at Indiana University, Bloomington, IN}}
}

\begin{abstract}
Using a high statistics sample of muon neutrino charged current quasielastic 
(CCQE) events, we report the first measurement of the double differential 
cross section ($\frac{d^2\sigma}{dT_\mu d\cos\theta_\mu}$) for this process.
The result features reduced model dependence and supplies the most complete 
information on neutrino CCQE scattering to date. Measurements of the absolute 
cross section as a function of neutrino energy ($\sigma[E_\nu^{QE,RFG}]$) and the single 
differential cross section ($\frac{d\sigma}{dQ^2_{QE}}$) are also provided, largely to 
facilitate comparison with prior measurements. This data is of particular use for 
understanding the axial-vector form factor of the nucleon as well as improving 
the simulation of low energy neutrino interactions on nuclear targets, which is 
of particular relevance for experiments searching for neutrino oscillations.
\end{abstract}

\maketitle

%%%%%%%%%%%%%%%%%%%%%%%%%%%%%%%%%%%%%%%%%%%%%%%%%%%%%%%%%
\section{CCQE event selection in MiniBooNE}
%%%%%%%%%%%%%%%%%%%%%%%%%%%%%%%%%%%%%%%%%%%%%%%%%%%%%%%%%
The MiniBooNE\footnote{The mini-Booster neutrino experiment (MiniBooNE)
at Fermi National Accelerator Laboratory (Fermilab)
is designed to search for 
$\nu_\mu \to \nu_e$ 
appearance neutrino oscillations~\cite{MB_osc}.} detector, a spherical tank filled with mineral oil, 
is surrounded by 1280 8'' photomultiplier tubes (PMTs) to detect 
\v{C}erenkov light from charged particles
\footnote{Detailed information about the Fermilab Booster neutrino beamline and 
the MiniBooNE neutrino detector are available elsewhere~\cite{MB_flux,MB_dtec}.}. 
In the 19.2$\mu s$ readout window, a ``subevent'' is defined as a timing cluster of PMT hits. 
The identification of $\numu$ CCQE interactions relies solely on 
the detection of the primary muon \v{C}erenkov light (first subevent) and 
the associated decay electron \v{C}erenkov light (second subevent) in 
these events~\cite{MB_CCQE}:
\beq
\begin{array}{cccl}
1: & \numu + n  & \to &  \mum + p \\ 
2: &            &     &  \mum \to \elm + \nueb + \numu.
\end{array}
\eeq
where each line in this equation identifies the subevent
where each process occurs.
Therefore, a CCQE candidate is characterized with a total of 2 subevents.  
After cuts, $\evki$ events are identified from 
$\MBosc1POT$ protons on target collected between 
August 2002 and December 2005. The cuts are estimated to be $\distCCQEef$\% efficient 
at selecting $\numu$ CCQE events in a 550 cm radius, with a CCQE purity 
of $\distCCQEpu$\%. 

The largest background is that from CC single-pion production (CC1$\pip$). 
The CC1$\pip$ interaction, proceeds as, 
\begin{equation}
\begin{array}{cccl}
1: & \numu +p(n) & \to & \mum + p(n) + \pip \; , \; \pip \to  \mup + \numu \\
2: &             &     & \mum \to \elm + \nueb + \numu \\
3: &             &     & \mup \to \elp + \nue  + \numub.
\end{array}
\end{equation}
Note this interaction results in total 3 subevents, 
the primary interaction and 2 muon decays resulting in an electron and a positron. 
Although these events can be removed from the CCQE sample by requiring 
only one muon decay (a total of 2 subevents), there is still a significant number of CC1$\pip$ events 
that contribute to the CCQE background because one of the muon decays may be missed for various reasons. 
Among them, $\pip$ absorption is a large effect (>40\%) with large uncertainty ($\sim$30\%).  
Additionally, the prediction of CC1$\pip$ backgrounds in the CCQE sample rely on 
the Rein and Sehgal's model~\cite{Rein-Sehgal} and final state interactions (FSIs) 
in the $\nuance$ event generator~\cite{nuance} which are not sufficiently accurate 
for a precise background prediction to measure the absolute CCQE cross section.  

%%%%%%%%%%%%%%%%%%%%%%%%%%%%%%%%%%%%%%%%%%%%%
\section{CC1$\pip$ background measurement}
%%%%%%%%%%%%%%%%%%%%%%%%%%%%%%%%%%%%%%%%%%%%%

Because of uncertainties in the CC1$\pip$ background predictions, 
we instead measure the CC1$\pip$ rate in our CC1$\pip$ data 
and the event generator is adjusted to match. 
By this, the predicted kinematic distribution of CC1$\pip$ events is modified, 
and the systematic error of CC1$\pip$ cross section is reduced to the level of the $\pip$ absorption uncertainty. 

\begin{figure}
\includegraphics[height=2.3in]{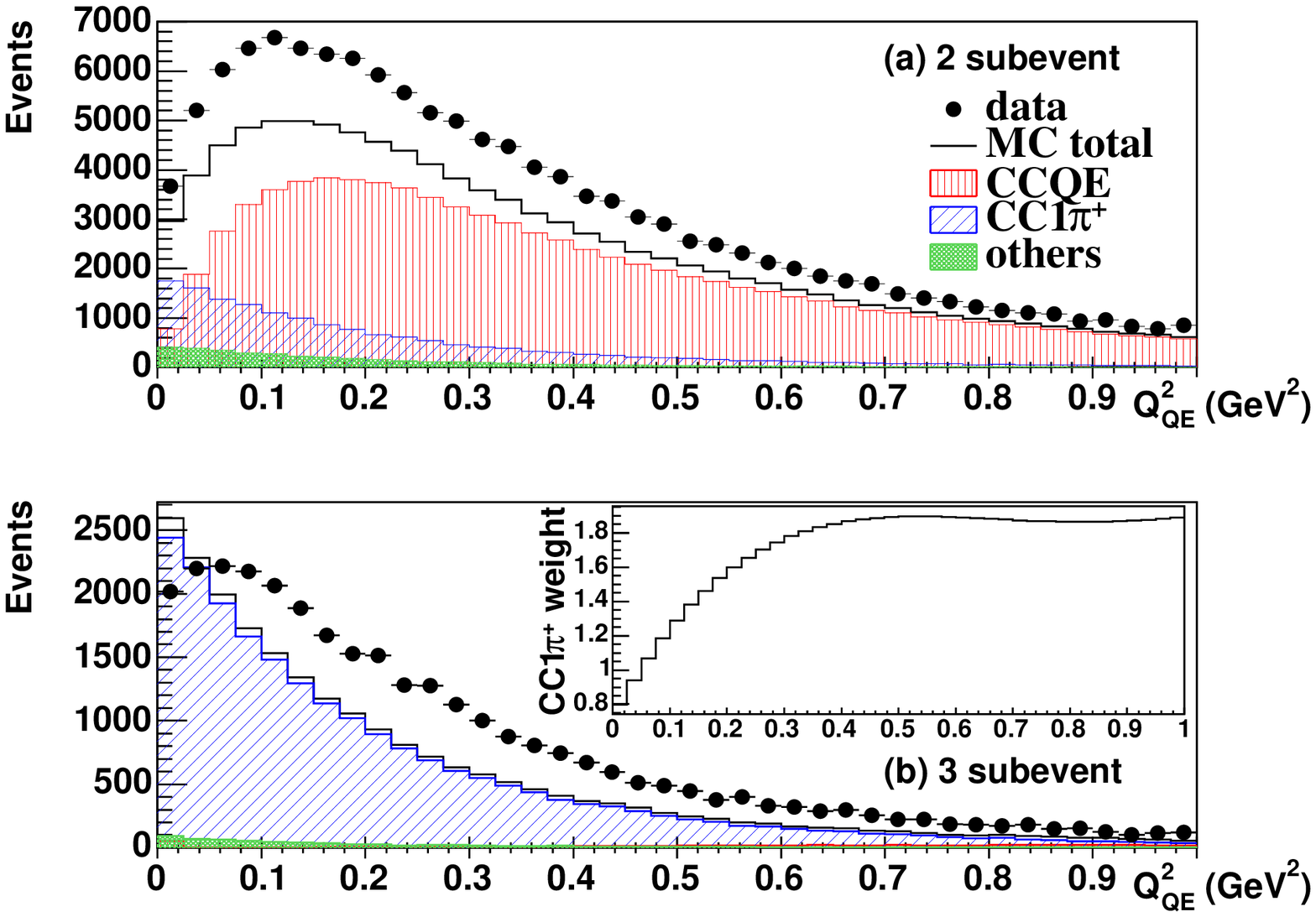}
\includegraphics[height=2.3in]{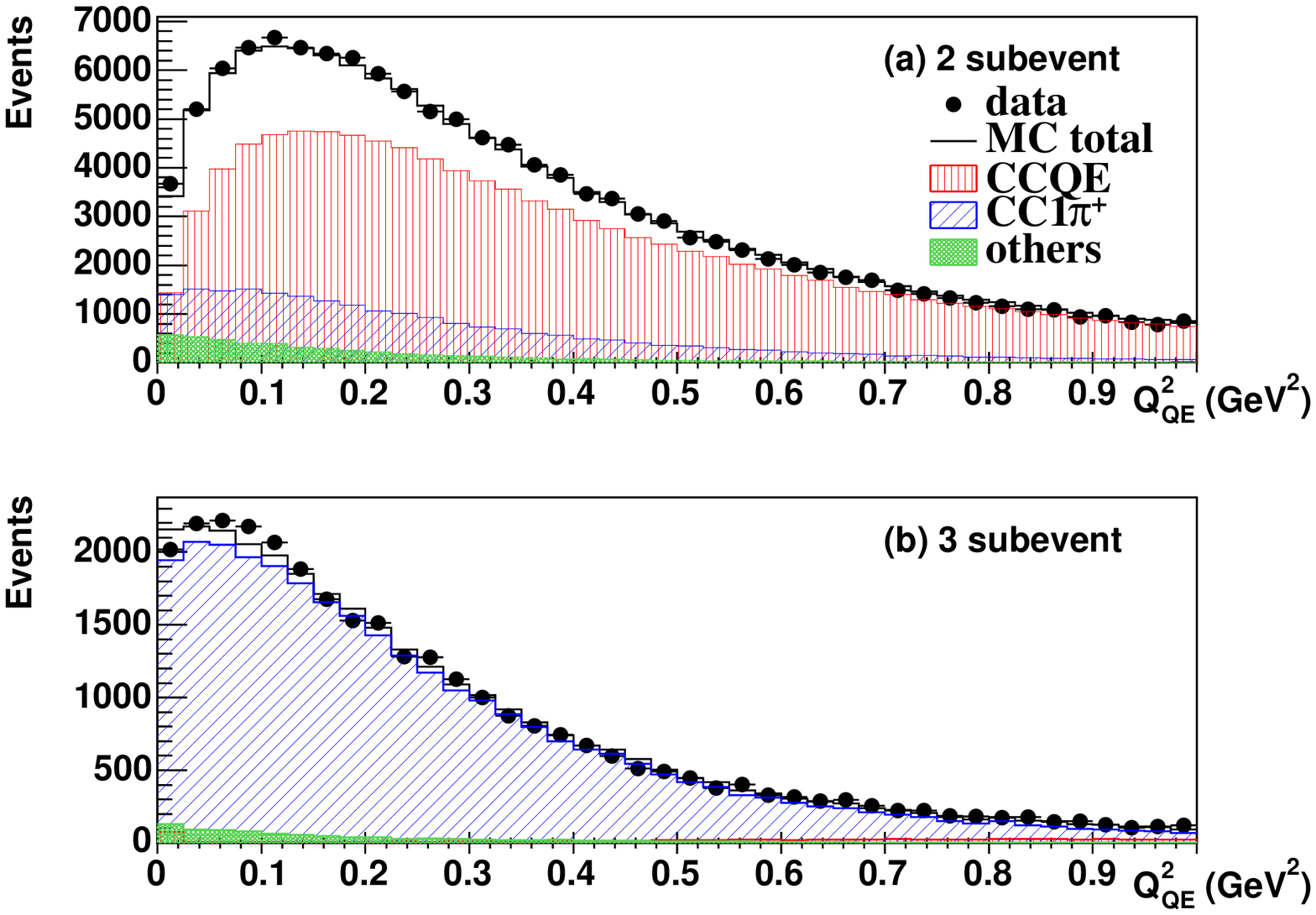}
\caption{
(color online). The distribution of events in $Q^2_{QE}$ for the (a) 2 and (b) 3 subevent samples. 
Data and MC samples are shown along with the individual MC contributions from 
CCQE, CC1$\pi^+$, and other channels. 
The left side is before the application of the CC1$\pi^+$ background correction. 
The inset in (b) shows the CC1$\pi^+$ reweighting function 
as determined from the background fit procedure. 
The right side is the same distribution after the application of the CC1$\pi^+$ background 
correction and the new CCQE model parameters $M_A^{eff}$ and $\ka$ 
as determined from the fit procedure described in the text.}
\label{fig:fit_before}
\end{figure}

The left plot in of Figure~\ref{fig:fit_before} shows the 
$Q^2_{QE}$ distributions\footnote{The neutrino energy $E_{\nu}^{QE}$ and 4-momentum transfer $Q^2_{QE}$ 
are reconstructed by assuming a CCQE interaction and neutron at rest, 
with averaged nucleon binding energy = 34~MeV.} 
for data and Monte Carlo (MC) of the two samples before the reweighting of 
CC1$\pi^+$ MC events.  The 2-subevent sample shows good shape agreement between data and MC. 
$\nuance$ uses the relativistic Fermi gas (RFG) model~\cite{Smith-Moniz} for CCQE interactions. 
In the previous work, we adjusted 2 parameters in RFG model, 
the effective axial mass $M_A^{eff}$ and Pauli blocking parameter $\ka$, 
to match the shape of the $Q^2_{QE}$ distribution to data~\cite{MB_CCQE}. 
Note that analysis did not consider the overall normalization of events.  
The 3-subevent sample shows a large data-MC disagreement in both shape and normalization.  
Using these samples, a simultaneous fit was performed for the shape 
and normalization of the 3-subevent sample,
and the normalization of the 2-subevent sample.  
These were then used to determine the CC1$\pi^+$ reweighting function 
which is shown in the inset plot of Figure~\ref{fig:fit_before}b (left). 
In order to reduce the sensitivity to the details of the shape of the 2-subevent sample,
only the 0.2$<Q^2_{QE}(\uGeVt)<$0.6 region was considered for the normalization parameter of this function.  
The $Q^2_{QE}$ shape of the CCQE sample was fit later 
although it has no impact on the cross section measurements. 
The  effect of the CCQE normalization on the 3-subevent sample was minimal since the background
from CCQE in this $Q^2_{QE}$ region is small as can be seen in the left plot of Figure~\ref{fig:fit_before}b. 
As a final step, with the measured CC1$\pi^+$ background incorporated, a shape-only fit 
to the 2-subevent (CCQE) sample is performed 
in order to extract revised CCQE model parameters~\cite{MB_CCQE}. 
The normalization of the CCQE sample is then extracted from the fit described above. 
The $Q^2_{QE}$ distributions of data from all subevent samples is shown
together with the MC prediction in the right plot of Figure~\ref{fig:fit_before}.  
Data-MC agreement is good in both subevent samples. 
A fit to the 2-subevent sample provided adjusted CCQE model parameters, $M_A^{eff}$ and $\ka$. 
This was a ``shape-only'' fit, that is, the MC was normalized with an arbitrary factor 
to have the same integrated event count as the background subtracted data.
The fit yielded,
\beq
M_A^{eff}&=& \NEWMAcon \pm \NEWMAerr~\uGeVct~;\no\\
\ka      &=& \NEWKAcon \pm \NEWKAerr        ~;\no\\
\ch^2/dof&=& \CHcon /38             ~.\no
\eeq

\begin{figure}
\includegraphics[height=2.3in]{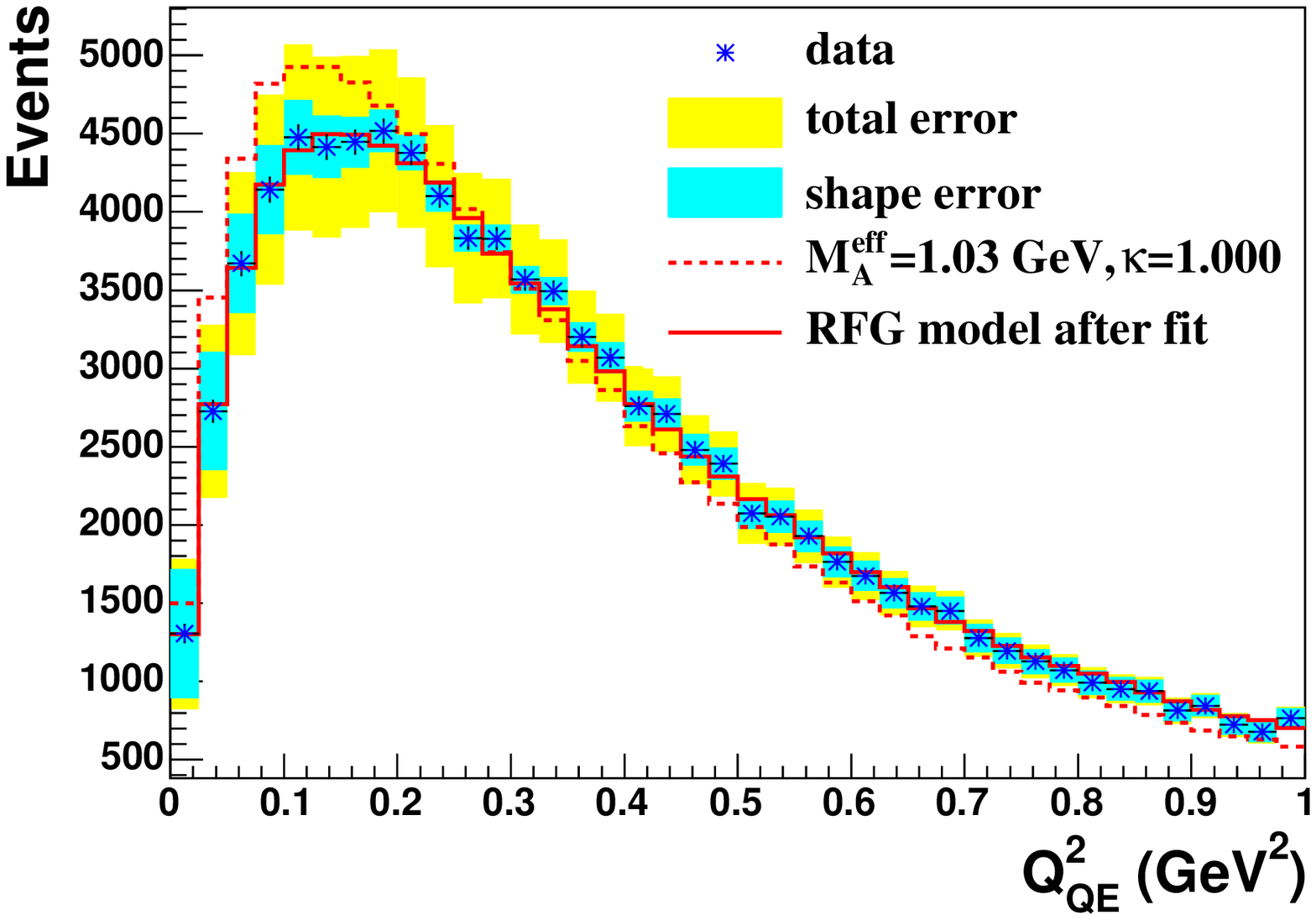}
\includegraphics[height=2.3in]{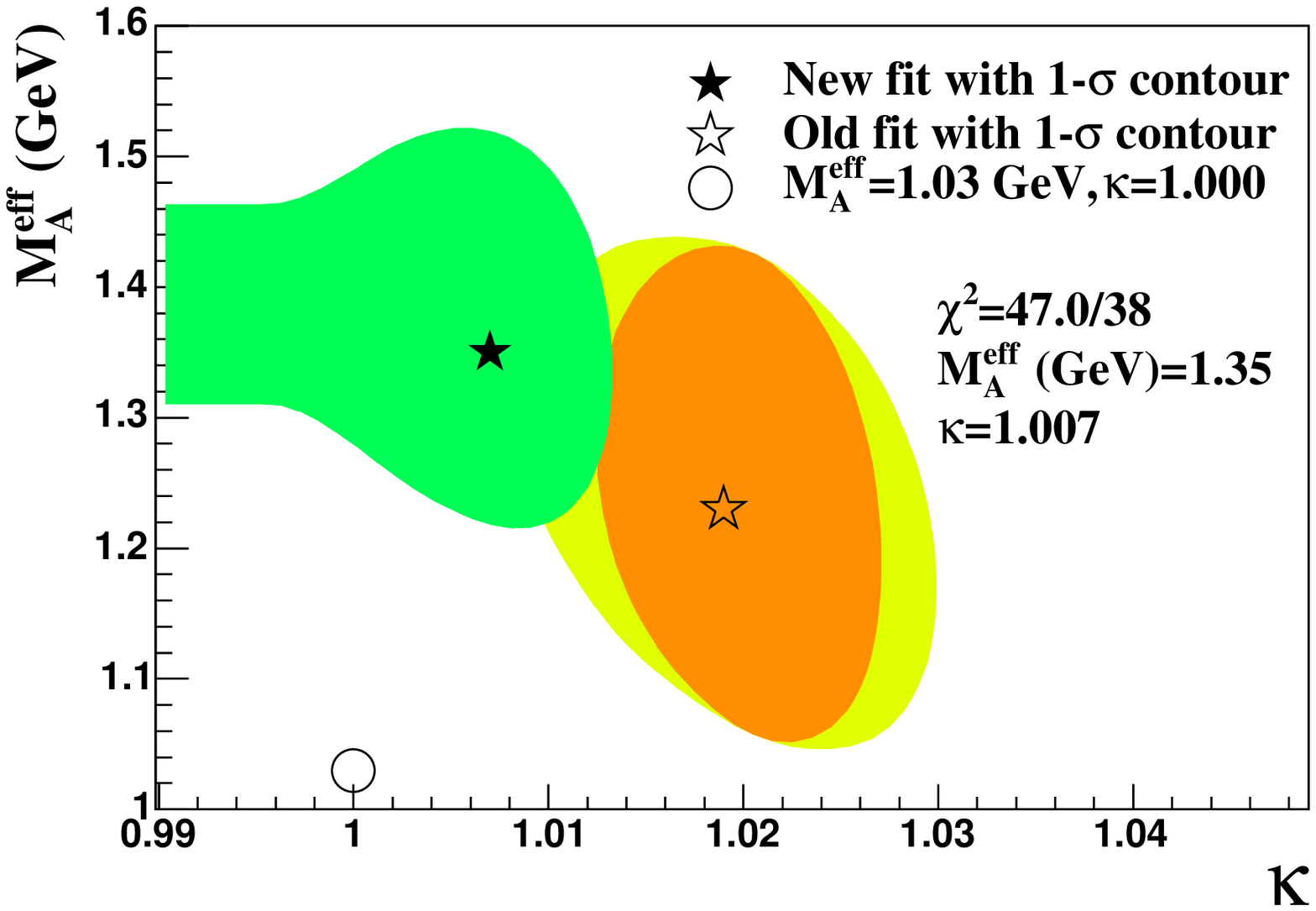}
\caption{\label{fig:contour}(Color online).  
Left plot is the $Q^2_{QE}$ distribution of the data, MC before, and MC after the fit with errors. 
Right plot is the $1-\si$ contour plot for the $M_A^{eff}-\ka$ fit. 
The filled star shows the best fit point and $1-\si$ contour 
extracted from this work. 
The open star indicates the best fit point and 
$1-\si$ contour from the previous work~\cite{MB_CCQE}.
Two regions are shown from the previous work, 
the larger area indicates the total uncertainty on the results 
including the background uncertainty~\cite{MB_CCQE}.}
\end{figure}

The left plot of Figure~\ref{fig:contour} shows the $Q^2_{QE}$ distribution of data, 
MC before, and MC after the fit with all sources of error. 
Data and MC after the fit agree within shape errors.   
The right plot of Fig.~\ref{fig:contour} is the $1-\si$ contour regions of this fit together
with the results from the previous MiniBooNE analysis~\cite{MB_CCQE}.  
Note that the current result is consistent (to within $1-\si$) with $\ka=1$.
This is because the CC1$\pi^+$ background resulting from the procedure
in this work has changed by an amount only just consistent with the error
assigned on the background in the previous work.  The value for $\ka$ is 
quite sensitive to the CC1$\pi^+$ background at low $Q^2_{QE}$. 
However, the previous and current results are consistent at the $1-\si$ level.

\begin{figure}
\includegraphics[height=2.3in]{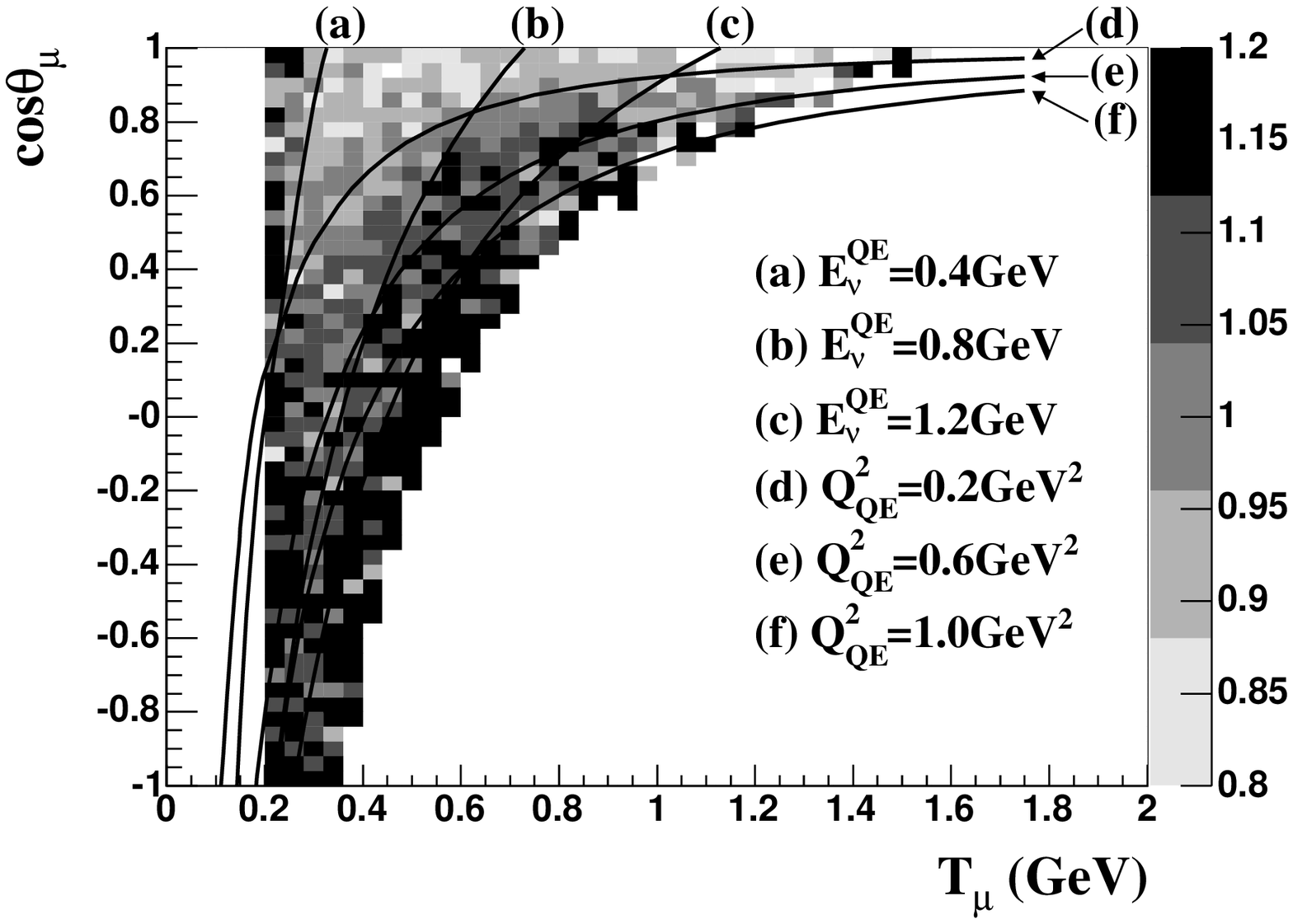}
\includegraphics[height=2.3in]{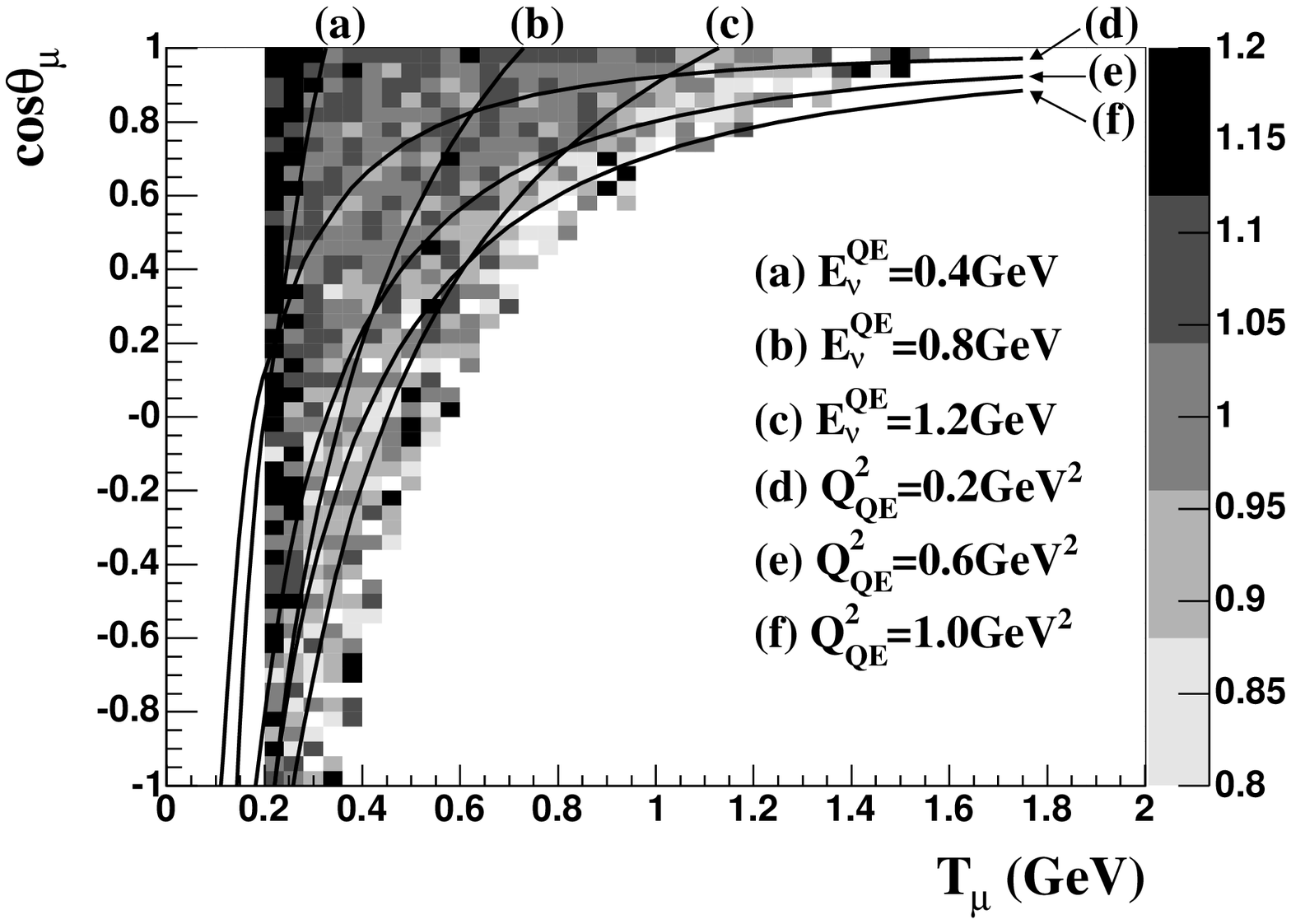}
\caption{Ratio of MiniBooNE $\numu$ CCQE data/simulation as a function of 
measured muon angle and kinetic energy. 
Left plot, with world averaged $M_A^{eff}$~(=$\NUAMAcon~\uGeVct$) and $\ka$~(=$\NUAKAcon$), 
and right plot, with newly determined $M_A^{eff}$~(=$\NEWMAcon~\uGeVct$) and $\ka$~(=$\NEWKAcon$). 
The ratio forms a 2D surface whose values are represented by the gray scale, 
shown on the right. If the simulation modeled the data perfectly, the ratio 
would be unity everywhere. Contours of constant $E_\nu$ and $Q^2$ are overlaid.}
\label{fig:2sub_2dim}
\end{figure}

The effect of the new $M_A^{eff}$ is clearly seen in 2-dimensional plots. 
Figure~\ref{fig:2sub_2dim} shows the data-MC ratio of CCQE candidate events as a function of 
muon kinetic energy $T_{\mu} (\uGeV)$ and muon scattering angle $cos\th_{\mu}$. 
Note the muon energy and muon scattering angle observables are the basis of 
all reconstructed kinematics variables in the $\numu$ CCQE channel in MiniBooNE.
In the left plot, we use the world averaged nuclear parameters 
($M_A^{eff}=\NUAMAcon~\uGeVct$, $\ka=\NUAKAcon$)~\cite{past-ma}. 
As can be seen, data-MC disagreement follows auxiliary lines of equal $Q^2$. 
This is the same tendency observed in the previous CCQE analysis in MiniBooNE~\cite{MB_CCQE}, 
indicating that data-MC disagreement is more likely due to an incorrect cross section prediction  
(=function of $Q^2$) than an incorrect flux prediction (=function of neutrino energy). 
After introducing the new $M_A^{eff}$ and $\ka$ ($M_A^{eff}=\NEWMAcon~\uGeVct$, $\ka=\NEWKAcon$), 
Fig.~\ref{fig:2sub_2dim} right plot, 
data-MC disagreement is reduced and we obtain an improved cross section prediction across the entire kinematic space. 

Note, this modification of the CCQE cross section prediction does not affect the CCQE absolute cross section measurement, 
presented below.   

%%%%%%%%%%%%%%%%%%%%%%%%%%%%%%%%%%%%%%%%%%%%%%%%%%%%%
\section{CCQE absolute cross section measurements}
%%%%%%%%%%%%%%%%%%%%%%%%%%%%%%%%%%%%%%%%%%%%%%%%%%%%%

%%%%%%%%%%%%%%%%%%%%%%%%%%%%%%%%%%%%%%%%%%%%%%%%%%%%%%%%%%%%%%%
\subsection{Flux-averaged double differential cross section}
%%%%%%%%%%%%%%%%%%%%%%%%%%%%%%%%%%%%%%%%%%%%%%%%%%%%%%%%%%%%%%%

\begin{figure}
\includegraphics[width=\columnwidth]{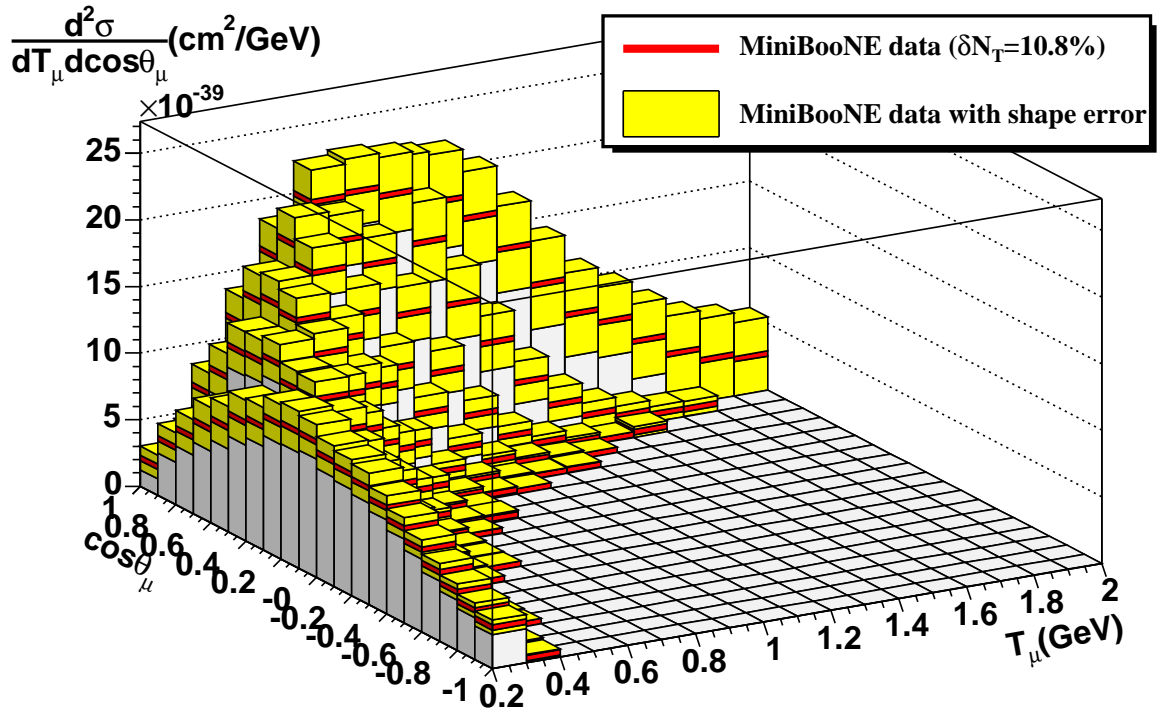}
\caption{\label{fig:ddsigma}(Color online).  
The flux-averaged double differential per nucleon ($n$) cross section for 
the $\nu_\mu$ CCQE process.  The dark bars indicate the measured values and
the surrounding lighter bands show the shape error.  The overall normalization 
(scale) error is 10.8\%.}
\end{figure}

Figure.~\ref{fig:ddsigma} shows the flux-averaged double differential cross section, 
$\frac{d^2\si}{dT_\mu dcos\th_\mu}$, for the $\nu_\mu$ CCQE process. 
The flux-averaged total cross section, an integral of the double differential cross section 
($-1<cos\th_\mu<+1$ and $0<T_\mu(\uGeV)<\infty$) is $\absXScon\times 10^{-39}$~$\ucmt$.
The total normalization error on this measurement is $\ttXSerr$\%.

The kinematic quantities, $T_\mu$ and $cos\th_\mu$, have been corrected for detector 
resolution effects only.  This result is the most model-independent
measurement of this process possible with the MiniBooNE detector. No cuts on the recoil
nucleons are used to define this process. The neutrino flux is an absolute prediction and
was not adjusted based on measured processes in the MiniBooNE detector.

%%%%%%%%%%%%%%%%%%%%%%%%%%%%%%%%%%%%%%%%%%%%%%%%%%%%%%%%
\subsection{Flux-averaged differential cross section}
%%%%%%%%%%%%%%%%%%%%%%%%%%%%%%%%%%%%%%%%%%%%%%%%%%%%%%%%

\begin{figure}
\includegraphics[height=3.0in]{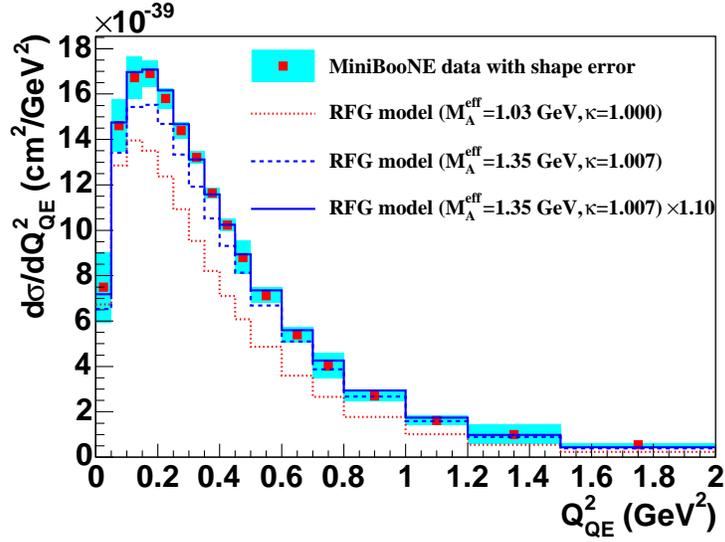}
\caption{\label{fig:dsigma}(Color online).  
The flux-averaged single differential per nucleon ($n$) cross section 
for the $\nu_\mu$ CCQE process. The measured values are shown as points 
with the shape error as shaded bars. Predictions from the $\nuance$ RFG
model with different values for the model parameters are shown as histograms.}
\end{figure}

Figure.~\ref{fig:dsigma} shows the flux-averaged single differential cross section, $\frac{d\si}{dQ^2_{QE}}$.  
The reconstructed 4-momentum transfer $Q^2_{QE}$ depends only upon the (unfolded) 
quantities $T_\mu$ and $cos\th_\mu$.  

In addition to the experimental result, Figure~\ref{fig:dsigma} also shows 
the prediction for the CCQE process from the $\nuance$ simulation with three
different variations of parameters in the underlying RFG model.  The predictions 
are flux-averaged and absolutely normalized.  
The RFG model is plotted with both the world-averaged CCQE parameters 
($M_A=\NUAMAcon~\uGeV$,$\ka=\NUAKAcon$) and with the CCQE parameters extracted 
from this analysis ($M_A=\NEWMAcon~\uGeV$, $\ka=\NEWKAcon$). The model
using the world-averaged CCQE parameters underpredicts the measured values 
significantly (by $\approx 30$\%). The model using the CCQE parameters extracted 
from the shape fit to the MiniBooNE CCQE data are within $\approx 10$\% of the data, 
consistent within the normalization uncertainty of $\approx 10$\%. 
The prediction with the CCQE parameters from
this analysis scaled by 1.10 is also plotted and is in good agreement with the data.

%%%%%%%%%%%%%%%%%%%%%%%%%%%%%%%%%%%%%%%%%%%%%%%%%
\subsection{Flux-unfolded total cross section}
%%%%%%%%%%%%%%%%%%%%%%%%%%%%%%%%%%%%%%%%%%%%%%%%%

\begin{figure}
\includegraphics[height=3.0in]{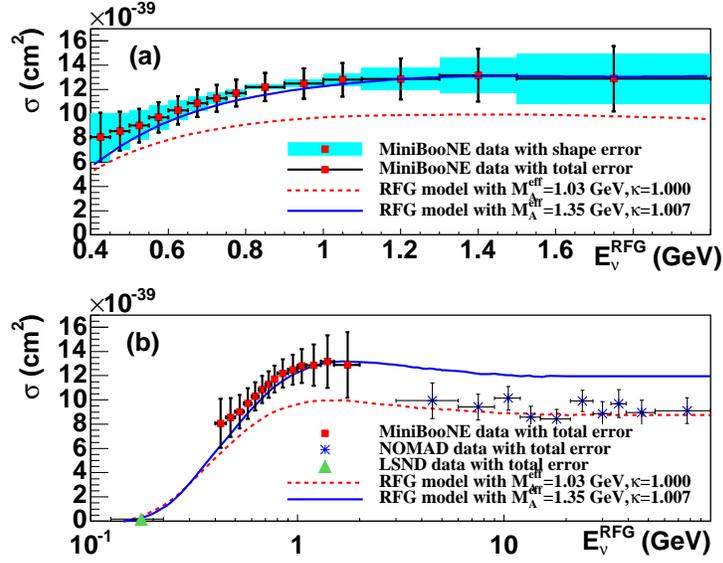}
\caption{\label{fig:sigma}(Color online).  
The flux-unfolded total per nucleon (n) cross section with total
errors and bin widths plotted indicated with the data points. In (a) shape errors 
are shown as shaded boxes. In (b), a larger energy
range is shown along with results from the LSND~\cite{LSNDxs}
and NOMAD~\cite{NOMAD} experiments.  Predictions from the $\nuance$ simulation
with two different RFG parameter variations are shown in both plots.}
\end{figure}

The flux-unfolded total cross section ($\si[E_\nu^{QE,RFG}]$) 
as a function of estimated neutrino energy $E_\nu^{QE,RFG}$ is shown in Figure~\ref{fig:sigma}.
The quantity $E_\nu^{QE,RFG}$ is a model-dependent estimate of the 
neutrino energy obtained after correcting for both detector and nuclear model
resolution effects.  These results depend on the details of the nuclear model
used for the calculation. The dependence is only weak in the peak of the flux
distribution but becomes strong at $E_\nu<0.5$~GeV and $E_\nu>1.0$~GeV,
in the ``tails'' of the flux distribution.  

In Figure~\ref{fig:sigma} data are compared with the $\nuance$ implementation
of the RFG model with the world averaged parameter values ($M_A^{eff}=\NUAMAcon~\uGeV$, $\ka=\NUAKAcon$), 
and the parameters extracted from this work ($M_A^{eff}=\NEWMAcon~\uGeV$, $\ka=\NEWKAcon$). 
These are absolute predictions from the model --- they are not scaled in any way.
The measurement is $\sim$20\% higher than the RFG model prediction with world
average parameter values at the flux peak ($700-800~\uMeV$).  The prediction
with the RFG parameter values extracted from the {\em shape-only}  fit to 
MiniBooNE CCQE data reproduces the data significantly better, 
to within $1\sigma$ for every point over the entire measured energy range.

Figure~\ref{fig:sigma}(b) shows the CCQE results from the LSND~\cite{LSNDxs} 
and NOMAD~\cite{NOMAD} experiments. It is interesting to note that NOMAD results
are better described with the world-average $M_A^{eff}$ and $\ka$ values. 

At this time, a solution to this growing mystery is not evident. 
Although there are tremendous efforts to model this process~\cite{new-model}, 
no models seem to be able to produce 
the (1) large observed $M_A^{eff}$ and (2) large observed total cross section, 
while keeping the ``bare'' $M_A=\NUAMAcon~\uGeV$ (the world averaged value). 
Model-independent cross section results from 
the MINOS near detector~\cite{MINOS}, running with $E_\nu\sim$3~GeV
and near-future experiments
such as MINERvA~\cite{MINERvA}, running with $2<E_\nu<20$~GeV could
help shed further light on this subject.

%%%%%%%%%%%%%%%%%%%%%%%%%%%%%%%%%%%%%%%%%%%%%%%%%%
%%%% BACKMATTER
%%%%%%%%%%%%%%%%%%%%%%%%%%%%%%%%%%%%%%%%%%%%%%%%%%
%%
%%\begin{theacknowledgments}
%%  Infandum, regina, iubes renovare dolorem, Troianas ut opes et
%%  lamentabile regnum cruerint Danai; quaeque ipse miserrima vidi, et
%%  quorum pars magna fui. Quis talia fando Myrmidonum Dolopumve aut duri
%%  miles Ulixi temperet a lacrimis?
%%\end{theacknowledgments}

%%%%%%%%%%%%%%%%%%%%%%%%%%%%%%%%%%%%%%%%%%%
%% The following lines show an example how to produce a bibliography
%% without the help of the BibTeX program. This could be used instead
%% of the above.
%%%%%%%%%%%%%%%%%%%%%%%%%%%%%%%%%%%%%%%%%%%

\end{document}